\PassOptionsToPackage{table}{xcolor}
\documentclass[letterpaper]{article}
\usepackage[preprint]{aaai2027}
\usepackage[hyphens]{url}
\usepackage{graphicx}
\usepackage{natbib}
\usepackage{caption}
\usepackage{latexsym}
\usepackage{array}
\usepackage{booktabs}
\usepackage{multirow}
\usepackage{amsmath}
\usepackage{amssymb}
\usepackage{tikz}
\usepackage{verbatim}
\usepackage[
  colorlinks=true,
  linkcolor=blue,
  citecolor=blue,
  urlcolor=blue,
  filecolor=blue
]{hyperref}

\definecolor{PromptBackground}{HTML}{F7F8FA}
\definecolor{PromptFrame}{HTML}{B8C2CC}
\definecolor{PromptTitleBackground}{HTML}{E9EEF5}
\definecolor{PromptTitleText}{HTML}{243447}
\definecolor{PromptText}{HTML}{263238}

\makeatletter
\newcommand{\promptverbatiminput}[1]{%
  \begingroup
  \def\verbatim@font{\ttfamily\footnotesize\color{PromptText}}%
  \verbatiminput{#1}%
  \endgroup}
\makeatother

\newcommand{\emptycircle}{\tikz[baseline=-0.55ex]\draw[line width=0.45pt] (0,0) circle (0.55ex);}
\newcommand{\halfcircle}{\tikz[baseline=-0.55ex]{\fill (0,-0.55ex) arc[start angle=-90,end angle=90,radius=0.55ex] -- cycle;\draw[line width=0.45pt] (0,0) circle (0.55ex);}}
\newcommand{\fullcircle}{\tikz[baseline=-0.55ex]\fill (0,0) circle (0.55ex);}
\newcommand{\circlednum}[1]{\textbf{(#1)}}
\newcommand{\blackcirclednum}[1]{%
  \tikz[baseline=(char.base)]{
    \node[circle,fill=black,text=white,inner sep=0.7pt,
          font=\bfseries\scriptsize] (char) {#1};}}
\DeclareRobustCommand{\toolname}{\textsc{SWE-RPG}}
\newcolumntype{L}[1]{>{\raggedright\arraybackslash}p{#1}}
\newcolumntype{C}[1]{>{\centering\arraybackslash}p{#1}}

\title{\toolname{}: A Unified Issue Resolution Benchmark for Requirement Clarification, Planning, and Code Generation for Coding Agents}

\author{
Xin Zhou\textsuperscript{\rm 1},
Chun Yong Chong\textsuperscript{\rm 2},
Kisub Kim\textsuperscript{\rm 3},
Yun Peng\textsuperscript{\rm 4},
Rui Shu\textsuperscript{\rm 5},\\
Zihan Wu\textsuperscript{\rm 6},
Xu Han \textsuperscript{\rm 7},
Guowen Yuan\textsuperscript{\rm 8},
Zeyang Zhuang\textsuperscript{\rm 4},
Jounghoon Kim\textsuperscript{\rm 7},\\
Jeongjin Ju\textsuperscript{\rm 3},
Seongmin Ju\textsuperscript{\rm 3},
Taein Yoon\textsuperscript{\rm 3},
David Lo\textsuperscript{\rm 1}
}
\affiliations{
\textsuperscript{\rm 1}Singapore Management University, Singapore\\
\textsuperscript{\rm 2}Monash University Malaysia, Malaysia\\
\textsuperscript{\rm 3}Daegu Gyeongbuk Institute of Science and Technology (DGIST), Republic of Korea\\
\textsuperscript{\rm 4}The Chinese University of Hong Kong, Hong Kong SAR, China\\
\textsuperscript{\rm 5}North Carolina State University, USA\\
\textsuperscript{\rm 6}City University of Hong Kong, Hong Kong SAR, China\\
\textsuperscript{\rm 7}The Hong Kong University of Science and Technology, Hong Kong SAR, China\\
\textsuperscript{\rm 8}The University of Hong Kong, Hong Kong SAR, China
}

\begin{document}
\maketitle

\begin{abstract}
Large language model-powered coding agents are increasingly used to modify
existing code repositories, for example, by adding features or fixing bugs.
Yet existing repository-level benchmarks typically evaluate only whether the final patch
passes tests.
Satisfying a user request requires a long chain of interdependent reasoning and
decisions: an agent must recover explicit and implicit requirements, formulate
a repository-grounded implementation plan, and translate it into correct code.
A pass/fail outcome cannot characterize how an unsuccessful trajectory diverges
from the requirements and implementation process needed for a correct patch.
To address this gap, we introduce \toolname{}, a repository-level benchmark
that combines executable patch evaluation with validated ground-truth references (GTs) for
\circlednum{1}~\textit{\underline{R}equirement Clarification} and
\circlednum{2}~\textit{Implementation \underline{P}lanning}. These intermediate
GTs support retrospective, GT-aligned diagnosis of complete coding-agent
trajectories across clarification, planning, code generation, and artifact
submission.
\toolname{} comprises 163 tasks from 31 Python and Java repositories, including
113 bug fixes and 50 feature additions.
We evaluate 3 coding agents, including Claude Code, Codex, and OpenCode, with 6 LLM backends, including Claude-Sonnet-5 and GPT-5.6-Terra.
Results show that the evaluated popular coding agents still struggle to implement user
requests in existing repositories, achieving an average resolved rate of only
31.5\% on \toolname{}.
Intermediate-GT diagnosis further identifies implicit requirement recovery as the
main bottleneck, accounting for 24.5\%--46.0\% of agent runs.
This result suggests implicit-requirement recovery as a key candidate direction for
improving coding agents.
The benchmark data and evaluation code are available at
\url{https://github.com/Xin-Zhou-smu/SWE-RPG-Bench}.
\end{abstract}

\section{Introduction}

Large language model-powered coding agents are reshaping software development
by automating repository-level maintenance and evolution, from bug fixing
\citep{yang2024swe-agent} to feature implementation~\citep{li2025feabench}.
Reliably translating users' raw requirements into correct changes in existing
codebases is therefore central to their practical deployment.
To assess this capability, repository-level issue resolution benchmarks have progressed from
SWE-bench~\citep{jimenez2024swebench} to recent efforts such as RACE-bench
~\citep{liu2026racebench} and Dialogue SWE-Bench~\citep{king2026dialogueswebench}.
These benchmarks provide valuable end-to-end insights by evaluating generated
patches with executable tests. 
However, producing a correct patch requires a long chain of decisions spanning
requirement clarification, implementation planning, and code generation.
A final patch verdict cannot identify the actual bottleneck or reveal clear
directions for improving coding agents.

\begin{figure}[t]
  \centering
  \includegraphics[width=1\linewidth]{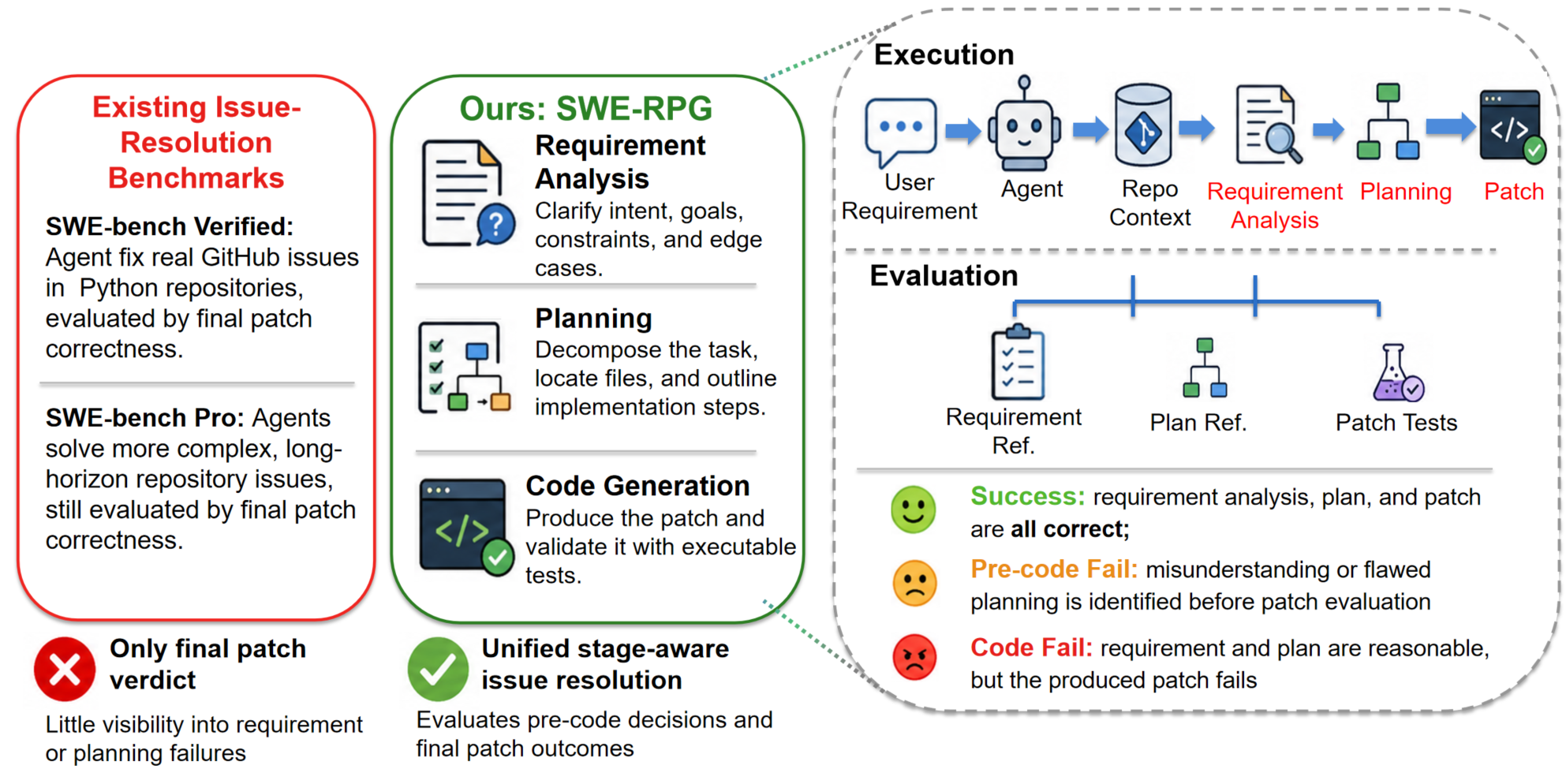}
  \caption{Comparison of \toolname{} with popular benchmarks. Unlike them,
  \toolname{} jointly evaluates implicit requirement
  clarification, planning, and code generation.}
  \label{fig:overview}
\end{figure}

\paragraph{Limitations of Existing Benchmarks.}
We argue that faithfully diagnosing coding agents requires observing the
complete issue-resolution\footnote{In repository-level issue-resolution task,
we call the original input request either the \emph{issue description} or the
\emph{user requirement}.} process.
Accordingly, we pose the following question: \textbf{When a coding agent fails,
does the primary bottleneck lie in
\blackcirclednum{1}~requirement clarification,
\blackcirclednum{2}~implementation planning, or
\blackcirclednum{3}~code generation?}
However, prior works cannot answer this question in three respects, as
summarized in Table~\ref{tab:stage-aware-comparison}:
\circlednum{1}~\textbf{Requirement Clarification.}
In practice, initial requirement descriptions are often incomplete or
ambiguous, requiring developers to recover implicit constraints from
task and repository context
\citep{knauss2015continuousclarification,franch2023statepractice}.
Recent benchmarks expand evaluation to long-horizon and continuous-evolution
tasks~\citep{deng2025swebenchproaiagents,thai2025sweevo,deng2025evoclaw},
assess high-level issue understanding~\citep{liu2026racebench}, or study
clarification in simulated dialogue~\citep{king2026dialogueswebench}.
However, none provides validated references for systematically evaluating
agents' recovery of implementation-critical requirements left implicit in
original requests.
\circlednum{2}~\textbf{Planning.}
Implementation plans translate intended behavior into actionable repository
changes~\citep{bairi2024codeplan} and should be sufficient to generate
functionally equivalent implementations. Most benchmarks provide no
implementation-plan references. RACE-bench comes closest with structured task
steps, but does not test their plan-to-code reproducibility.
\circlednum{3}~\textbf{Evaluation.}
Existing benchmarks primarily determine whether submitted patches pass
executable tests, revealing whether agents fail but not where. Some provide
module- or failure-level analyses, but none consistently attributes unsuccessful
trials to aligned requirement, planning, or coding stages.

\begin{table}[t]
  \centering
  \footnotesize
  \resizebox{\columnwidth}{!}{%
  \begin{tabular}{@{}lccccccc@{}}
  \toprule
  \textbf{Benchmark} & \textbf{Time} &
  \multicolumn{2}{c}{\circlednum{1} \textbf{Req. Clarification}} &
  \multicolumn{2}{c}{\circlednum{2} \textbf{Planning}} &
  \multicolumn{2}{c}{\circlednum{3} \textbf{Evaluation}} \\
  \cmidrule(lr){3-4}\cmidrule(lr){5-6}\cmidrule(lr){7-8}
  & \textbf{(yymm)} & \textbf{Implicit Req.} & \textbf{Categorized Ref.} &
  \textbf{Step Ref.} & \textbf{Reprod. Impl.} &
  \textbf{Exec. Patch} & \textbf{Stage Attr.} \\
  \midrule
  SWE-bench~\citep{jimenez2024swebench} & 2310 & \emptycircle & \emptycircle & \emptycircle & \emptycircle & \fullcircle & \emptycircle \\
  SWE-bench Verified~\citep{swebench-verified} & 2408 & \emptycircle & \emptycircle & \emptycircle & \emptycircle & \fullcircle & \emptycircle \\
  SWE-bench Multimodal~\citep{yang2025swebenchmultimodal} & 2410 & \emptycircle & \emptycircle & \emptycircle & \emptycircle & \fullcircle & \emptycircle \\
  Multi-SWE-bench~\citep{zan2025multi} & 2504 & \emptycircle & \emptycircle & \emptycircle & \emptycircle & \fullcircle & \emptycircle \\
  SWE-PolyBench~\citep{rashid2025swepolybench} & 2504 & \emptycircle & \emptycircle & \emptycircle & \emptycircle & \fullcircle & \emptycircle \\
  SWE-rebench~\citep{badertdinov2025swerebench} & 2505 & \emptycircle & \emptycircle & \emptycircle & \emptycircle & \fullcircle & \emptycircle \\
  SWE-bench-Live~\citep{zhang2025swebenchgoeslive} & 2505 & \emptycircle & \emptycircle & \emptycircle & \emptycircle & \fullcircle & \emptycircle \\
  SWE-bench Pro~\citep{deng2025swebenchproaiagents} & 2509 & \emptycircle & \emptycircle & \emptycircle & \emptycircle & \fullcircle & \halfcircle \\
  SWE-Compass~\citep{xu2025swecompass} & 2511 & \emptycircle & \emptycircle & \emptycircle & \emptycircle & \fullcircle & \halfcircle \\
  SWE-Evo~\citep{thai2025sweevo} & 2512 & \emptycircle & \emptycircle & \emptycircle & \emptycircle & \fullcircle & \halfcircle \\
  EvoClaw~\citep{deng2025evoclaw} & 2603 & \emptycircle & \emptycircle & \emptycircle & \emptycircle & \fullcircle & \halfcircle \\
  RACE-bench~\citep{liu2026racebench} & 2603 & \emptycircle & \emptycircle & \fullcircle & \emptycircle & \fullcircle & \halfcircle \\
  SWE-Explore~\citep{zhang2026sweexplore} & 2606 & \emptycircle & \emptycircle & \emptycircle & \emptycircle & \halfcircle & \emptycircle \\
  Dialogue SWE-Bench~\citep{king2026dialogueswebench} & 2606 & \halfcircle & \emptycircle & \emptycircle & \emptycircle & \fullcircle & \emptycircle \\
  \midrule
  \toolname{} (Ours) & -- & \fullcircle & \fullcircle & \fullcircle & \fullcircle & \fullcircle & \fullcircle \\
  \bottomrule
  \end{tabular}%
  }
  \caption{
  Comparison of \toolname{} with previous benchmarks.
  \toolname{} combines implicit requirements,
  implementation-sufficient planning, and stage-aware evaluation.
  \circlednum{1}~\textbf{Req. Clarification}: \textit{Implicit Req.} recovers
  implicit requirements; \textit{Categorized Ref.} provides
  gold references under a practitioner-informed taxonomy.
  \circlednum{2}~\textbf{Planning}: \textit{Step Ref.} provides actionable gold
  steps; \textit{Reprod. Impl.} verifies the plans can generate equivalent code.
  \circlednum{3}~\textbf{Evaluation}: \textit{Exec. Patch} evaluates patches
  with executable tests; \textit{Stage Attr.} assigns main failure reasons to
  specific stages.
  \protect\emptycircle/\protect\halfcircle/\protect\fullcircle denote
  no/partial/direct support.}
  \label{tab:stage-aware-comparison}
  \end{table}

\paragraph{Our Solution.}
We propose \toolname{}, a new stage-aware benchmark for evaluating coding
agents' capabilities in requirement clarification, planning, and coding during
repository-level issue resolution. \toolname{} addresses the aforementioned
limitations through three main features:
\circlednum{1}~\textbf{Requirement Clarification.}
\toolname{} evaluates agents' ability to recover implicit requirements using
gold clarification references across six practitioner-informed dimensions
(e.g., functional intent and technical context).
\circlednum{2}~\textbf{Planning.}
Each task provides implementation-sufficient gold plan steps, which we use to
assess the quality of implementation plans generated by coding agents.
\circlednum{3}~\textbf{Evaluation.}
\toolname{} evaluates final patches with executable tests, analyzes trajectories
to assess clarification and planning, and attributes failures to requirement,
planning, or coding stages.

\paragraph{Evaluation.}
We evaluate 3 coding-agent frameworks paired with 6 LLM backends. Agents achieve an average resolved rate of
only 31.5\%, with 46.7\% of agent runs failing primarily
during requirement clarification or implementation planning. Claude Code,
Codex, and OpenCode also expose distinct stage-aware failure profiles across
requirement, planning, implementation, and artifact stages. These findings
highlight distinct capability bottlenecks and show how stage-aware evaluation
can suggest candidate areas for improvement.

In summary, our contributions are as follows:
\begin{itemize}
  \item We propose \toolname{}, a stage-aware benchmark comprising 163 tasks
  from 31 Python/Java repositories. To our knowledge, it is the first
  repository-level benchmark to jointly evaluate agents' capabilities in
  requirement clarification, implementation planning, and code generation.

  \item We evaluate three popular coding-agent frameworks paired with six
  recent LLMs in a \(3\times6\) matrix on \toolname{}. Agents achieve an average resolved rate of only
  31.5\%, while 46.7\% of all agent runs fail during requirement
  clarification or implementation planning. Our stage-aware analyses reveal
  framework-specific bottlenecks and suggest candidate areas for improving
  coding agents.
\end{itemize}
\section{Related Work}

\paragraph{Repository-level Issue-resolution Benchmarks.}
SWE-bench established an executable benchmark for repository-level issue
resolution: agents modify code repositories to complete requirements, and
their patches are evaluated with tests~\citep{jimenez2024swebench}.
Later benchmarks improve reliability and freshness
~\citep{swebench-verified,badertdinov2025swerebench,zhang2025swebenchgoeslive}
or broaden programming-language and task coverage
~\citep{yang2025swebenchmultimodal,zan2025multi,rashid2025swepolybench,
deng2025swebenchproaiagents,xu2025swecompass,thai2025sweevo,deng2025evoclaw}.
However, as Table~\ref{tab:stage-aware-comparison} shows, most benchmarks 
primarily evaluate final patches and lack validated requirement and planning
references for diagnosing unsuccessful trajectories.

\begin{figure*}[t]
  \centering
  \includegraphics[width=\textwidth]{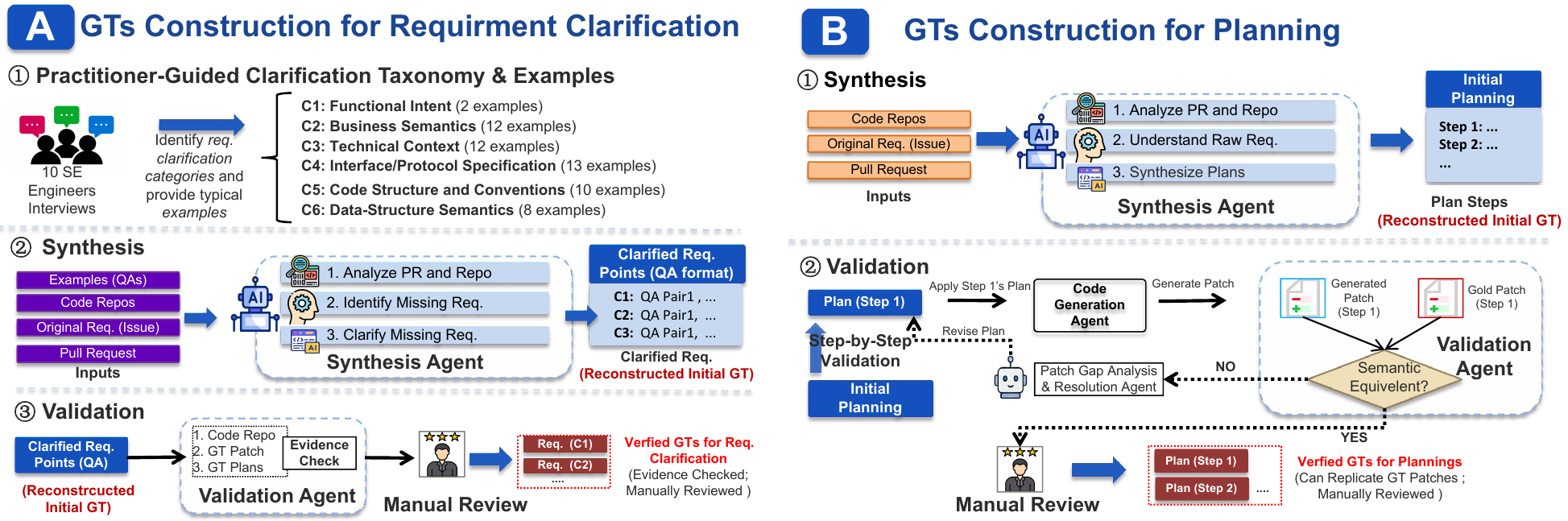}
    \caption{Ground-truths Construction for Requirement Clarification
    and Planning. Using the original requirement, repository, merged PR, and
    tests, we synthesize Clarification GT and an
    actionable implementation plan. Clarification GT is validated for
    evidential support, and cross-stage consistency. 
    Planning GT is validated step by step by checking whether each step
    can guide an implementation functionally equivalent to its gold subpatch.
    Finally, two authors independently review every resulting reference for
    evidential support and cross-stage consistency.}
  \label{fig:gt-construction}
\end{figure*}

\paragraph{Close Works.}
RACE-bench~\citep{liu2026racebench} and Dialogue
SWE-Bench~\citep{king2026dialogueswebench} are the two closest works.

\paragraph{\normalfont\itshape RACE-bench.} RACE-bench is a concurrent work that shares our
goal of exposing intermediate reasoning~\citep{liu2026racebench}. However, the
following two features distinguish \toolname{} from RACE-bench. First,
RACE-bench does not provide references for implicit requirements, a central challenge in
requirement clarification; it summarizes only the high-level intent already
stated in the original requirement. In contrast, \toolname{} provides
evidence-supported references for implementation-critical information not fully
specified in the initial request. Second, RACE-bench provides plan steps and validates their
semantic necessity, but does not test whether code generated from them can
reproduce a functionally equivalent code. In contrast, \toolname{}
directly checks plan-to-code reproducibility through code generation.

\paragraph{\normalfont\itshape Dialogue SWE-Bench.} Dialogue SWE-Bench reformulates
SWE-bench Verified~\citep{swebench-verified} tasks as interactive sessions. For
each task, it constructs a shortened initial query by deliberately omitting
critical details, then studies whether agents can recover them through simulated
dialogue. The resulting clarification setting is therefore synthetic and do not 
provide references mined from naturally implicit requirements. In
contrast, \toolname{} retains the original request and mines
implementation-critical implicit requirements from real development evidence.
We also provide reproducible planning references, which Dialogue SWE-Bench
lacks.

\section{\toolname{} Benchmark}
\label{sec:framework}

\toolname{} is a repository-level issue-resolution benchmark with executable
patch oracles and validated intermediate GTs.
In this section, we first formulate the benchmark task and its GT-aligned
diagnostic protocol,
then describe task and annotation construction and summarize the resulting
dataset.
Figure~\ref{fig:gt-construction} summarizes how we reconstruct and validate the
ground-truth references for Req. Clarification and Planning.

\subsection{Task Formulation and Example}
\label{sec:task-formulation}

\paragraph{Input and Output.}
The input to each task consists of (i) a repository checked out at a specified
base commit and (ii) a user requirement in natural language. Given
these inputs, the coding agent autonomously modifies the repository and
outputs a code patch.

\paragraph{Example Instance.}
Figure~\ref{fig:swerpg-example-data} illustrates a \toolname{} instance, in which TSV import incorrectly removes boundary whitespace
when \texttt{trimStrings} is disabled. The instance provides stage-specific
ground-truth artifacts for diagnosis:
\textit{Clarification GT} makes implicit requirements explicit, including the TSV-only
scope and unchanged CSV behavior; \textit{Plan GT} provides the implementation steps. 
The Gold Patch, Functional Tests, and Docker
environment support executable, reproducible patch evaluation. 
Together, these artifacts support stage-aware analysis and evaluation.

\begin{figure}[t]
  \centering
  \fcolorbox{blue!40!black}{blue!3}{%
    \begin{minipage}{\dimexpr\columnwidth-2\fboxsep-2\fboxrule\relax}
      \footnotesize
      \raggedright
      \color{black!90}

      {\colorbox{blue!65!black}{%
        \parbox{\dimexpr\linewidth-2\fboxsep\relax}{%
          \color{white}\textbf{Example: OpenRefine-6609.}}}}

      \textcolor{blue!65!black}{\textbf{Raw requirement.}}
      TSV import incorrectly removes boundary whitespace when
      \texttt{trimStrings} is disabled.

      \textcolor{blue!65!black}{\textbf{Clarification GT.}}
      \textbf{Functional Intent  (C1)}\\
      \emph{Q:} Regarding the fixing scope, should the fix change whitespace handling for all
      separator-based imports?\\
      \emph{A:} No. It must affect only the TSV path; CSV behavior must remain
      unchanged
      \emph{\textbf{\ldots}}

      \textcolor{blue!65!black}{\textbf{Plan GT.}}
      \textbf{Step 1:} Locate the TSV branch in
      \texttt{SeparatorBasedImporter.parseOneFile}.\\
      \textbf{Step 2:} Disable Univocity's automatic removal of leading and
      trailing whitespace in \texttt{TsvParserSettings} \textbf{\ldots}

      \textcolor{blue!65!black}{\textbf{Patch GT.}}
      {\ttfamily\scriptsize
        \textcolor{black!65}{TsvParserSettings settings = \ldots;}\\
        \textcolor{green!45!black}{+ settings.setIgnoreLeadingWhitespaces(false);}\\
        \textcolor{green!45!black}{+ settings.setIgnoreTrailingWhitespaces(false);}\\
        \textcolor{black!45}{\textbf{\ldots}}
        \par}

      \textcolor{blue!65!black}{\textbf{Tests.}}{ \ttfamily\scriptsize
      readDoesNotTrimLeadingTrailingWhitespace \textbf{\ldots}}

      \textcolor{blue!65!black}{\textbf{Environment:}}
      in a reproducible docker container
    \end{minipage}%
  }
  \caption{An illustrative \toolname{} instance.}
  \label{fig:swerpg-example-data}
\end{figure}

\subsection{Benchmark Construction}
\label{sec:construction}

\begin{table*}[t]
  \centering
  \small
  \begin{tabular*}{\textwidth}{@{\extracolsep{\fill}}ccccc@{}}
      \toprule
      \textbf{Repo and Task Collection} &
      \textbf{Task Filtering} &
      \textbf{Environment Build} &
      \textbf{Test Validation} &
      \textbf{Intermediate GT Validation} \\
      \midrule
      \shortstack{\textbf{2,000+}\\Eligible PR-Issue Pairs} &
      \shortstack{\textbf{1,000+}\\Relevant tasks} &
      \shortstack{\textbf{400+}\\With Runnable Environments} &
      \shortstack{\textbf{200+}\\With Stable Tests} &
      \shortstack{\textbf{163}\\With Verified Intermediate GTs} \\
      \bottomrule
  \end{tabular*}
  \caption{Benchmark Candidate reduction during benchmark construction.}
  \label{tab:construction-funnel}
\end{table*}

\paragraph{Repository and Task Selection.}
\toolname{} collects merged issue--PR pairs from mature, actively maintained
Python and Java repositories. Each pair must include reproducible tests, involve
a recent and traceable change, and not overlap with existing issue-resolution
benchmarks.
This screening yields 2000+ issue--PR instances with task-relevant tests from
100+ repositories.
We further retain only bug-fixing and new-feature tasks whose tests directly
exercise the target behavior, resulting in 1000+ candidate instances.

\paragraph{Building Environments.}
\label{sec:environments}
We reuse the automated agentic build pipeline of SWE-bench-Live~\citep{zhang2025swebenchgoeslive} to construct an
isolated Docker image for each candidate at its base commit. We extend this
pipeline with expert-guided recovery: when an automated build fails, the first author
inspects its logs, identifies missing dependencies or configuration, and feeds
these corrections into the next build attempt. We
validate each image by comparing test outcomes before and after applying the
gold patch. We repeat tests to remove flaky instances and manually verify that
fail-to-pass tests directly target the intended behavior. This process yields
400+ runnable images and 200+ qualified instances.

\paragraph{Overview of Ground-Truth Construction.}
We construct ground truth differently for the three stages. Original
requirements often omit implementation-critical details
\citep{bo2024chatbr,knauss2015continuousclarification,franch2023statepractice},
while existing benchmarks generally provide no explicit implementation-plan
references (Table~\ref{tab:stage-aware-comparison}). We therefore construct the
Requirement Clarification and Planning references through \textit{synthesis
and validation}. During synthesis, we derive the references
from the base repository, issue and PR discussion, developer patch, and selected
tests. During validation, we check them against the development evidence and
implemented behavior. We instantiate the agentic components of this construction
pipeline with GPT-5.4~\citep{openai2026gpt54}, the latest available model at the time of benchmark
construction, to maximize the quality of the synthesized references. For Code
Generation, no synthesis is needed: the merged
PR provides the developer patch, and its task-relevant tests provide the
executable correctness oracle. We detail the synthesis and validation procedure
for each stage below.

\begin{table}[!t]
  \centering
  \footnotesize
  \resizebox{\columnwidth}{!}{%
  \begin{tabular}{@{}lr@{\hspace{9pt}}lr@{}}
  \toprule
  \multicolumn{4}{c}{\textbf{\toolname{} Dataset Composition}} \\
  Tasks & 163 & Repositories & 31 \\
  Java / Python tasks & 85 / 78 & Java / Python repos. & 17 / 14 \\
  Bug fixes & 113 (69.3\%) & Feature implementations & 50 (30.7\%) \\
  \midrule
  \multicolumn{2}{c}{\textbf{Task Statistics}} &
  \multicolumn{2}{c}{\textbf{Plan GT}} \\
  \textbf{\textit{Statistic}} & \textbf{\textit{Avg. / Max.}} &
  \textbf{\textit{Statistic}} & \textbf{\textit{Avg. / Max.}} \\
  Requirement Words & 301.10 / 2,195 & Plan steps & 2.06 / 10 \\
  Codebase LOC & 272.51K / 1.59M & File locations & 2.47 / 10 \\
  Gold-patch LOC & 56.74 / 639 & Rule constraints & 11.80 / 50 \\
  Test cases & 2,248.56 / 15,119 & & \\
  \midrule
  \multicolumn{4}{c}{\textbf{Clarification GT}} \\
  \textbf{\textit{Statistic}} & \textbf{\textit{Avg. / Max.}} &
  \textbf{\textit{Statistic}} & \textbf{\textit{Avg. / Max.}} \\
  \textit{\# of Implicit Clarification Points} & \textit{4.53 / 8} & & \\
  C1 Functional intent & 1.21 / 4 & C4 Interface/protocol & 0.99 / 2 \\
  C2 Business semantics & 0.60 / 3 & C5 Structure/conventions & 0.47 / 1 \\
  C3 Technical context & 0.72 / 2 & C6 Data-structure semantics & 0.54 / 2 \\
  \bottomrule
  \end{tabular}%
  }
  \caption{Statistics of \toolname{} and its intermediate GTs.}
  \label{tab:benchmark-gt-statistics}
  \end{table}

  \begin{figure*}[!t]
    \centering
    \begin{minipage}[t]{0.49\textwidth}
      \centering
      \includegraphics[width=\linewidth]{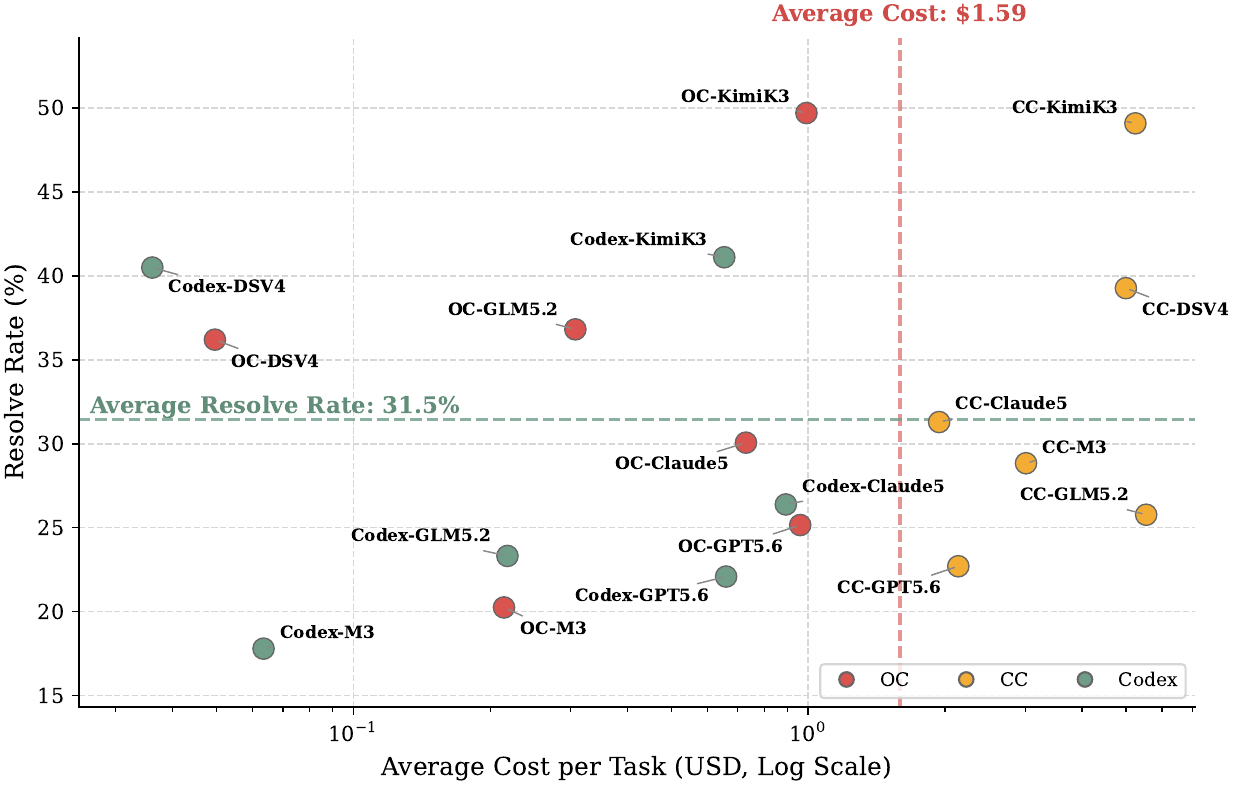}
      \textbf{(a) Cost versus resolve rate}
    \end{minipage}\hfill
    \begin{minipage}[t]{0.49\textwidth}
      \centering
      \includegraphics[width=\linewidth]{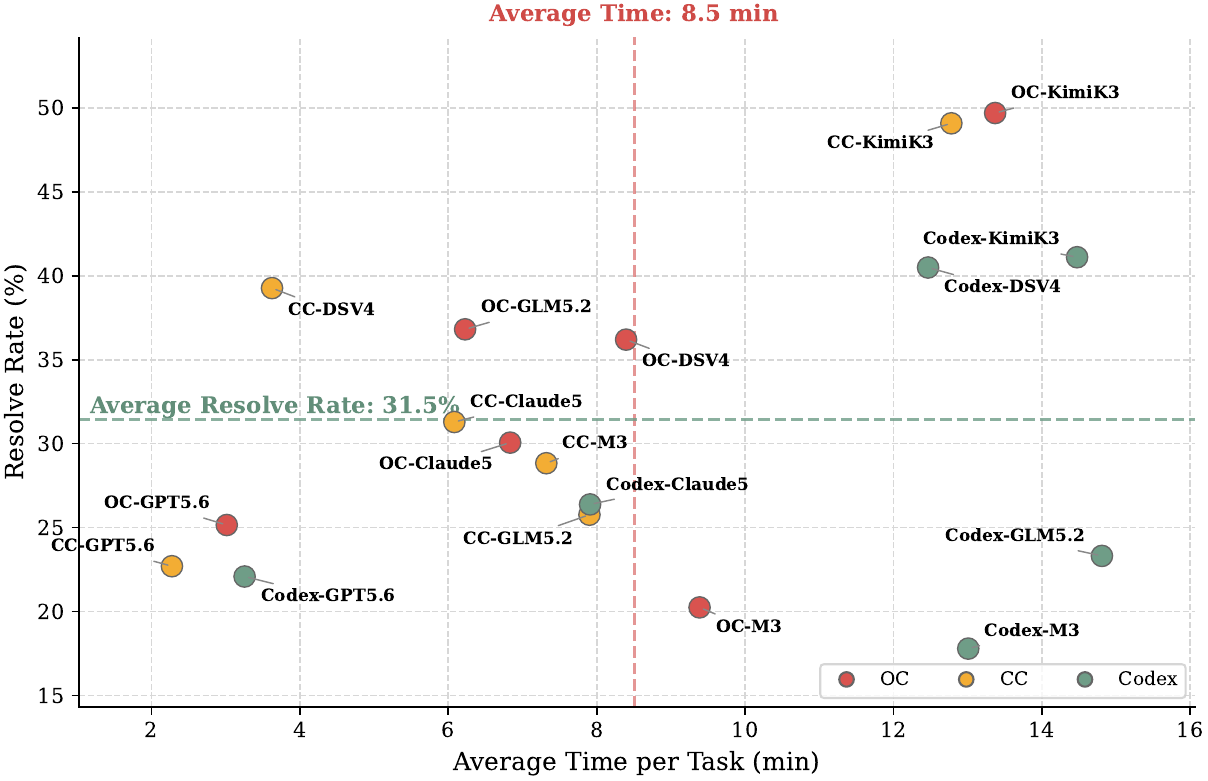}
      \textbf{(b) Time versus resolve rate}
    \end{minipage}
    \caption{Cost and execution time versus resolve rate.}
    \label{fig:cost-time-resolve}
  \end{figure*}
  
  \begin{figure}[!t]
  \centering
  \includegraphics[width=\columnwidth]{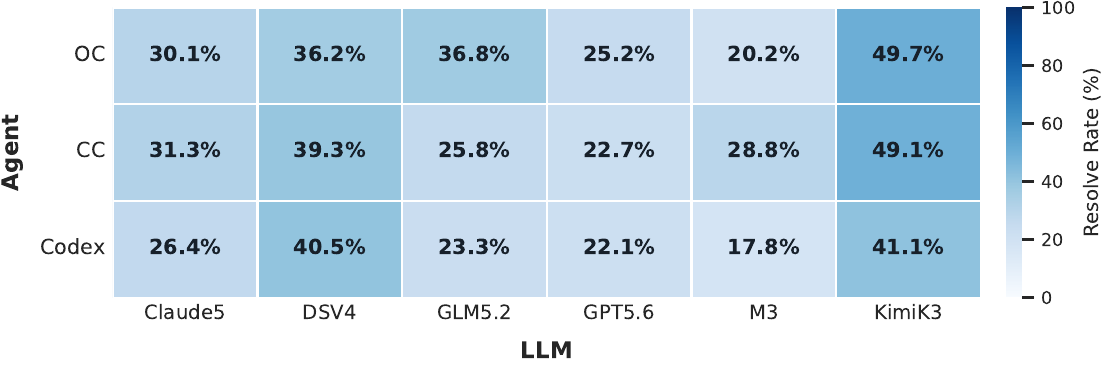}
  \caption{Resolve rates across agent--LLM combinations.}
  \label{fig:combo-heatmap}
  \end{figure}

\paragraph{Ground-Truth Construction for Req. Clarification.}
\paragraph{\normalfont\itshape Practitioner-Guided Clarification Taxonomy.}
Requirement clarification is difficult to define because it targets
implementation-critical requirements that remain implicit in the original
request. Which implicit decisions matter to developers, and which aspects they
repeatedly clarify in real software development, are unknown to some extent. 
We therefore conducted seminar-style interviews with ten experienced software
engineers from Fortune Global 500 technology companies. We asked what
information developers need before implementation, what missing information
they recover during clarification, and which omissions most often cause
incorrect implementations. Their responses served two purposes: deriving a
practitioner-grounded taxonomy and collecting representative QA examples for
GT synthesis. The resulting taxonomy uses standardized QA pairs and comprises
six categories: \textbf{\textit{C1: Functional Intent}},
\textbf{\textit{C2: Business Semantics}},
\textbf{\textit{C3: Technical Context}},
\textbf{\textit{C4: Interface and Protocol Specifications}},
\textbf{\textit{C5: Code Structure and Naming Conventions}}, and
\textbf{\textit{C6: Detailed Data-Structure Semantics}}. We curated 57 seed question-answer pairs.

\paragraph{\normalfont\itshape Synthesis and Validation.}
As illustrated in Figure~\ref{fig:gt-construction}, the taxonomy guides a
\textit{synthesis-and-validation} pipeline. Using the corresponding seed QA examples as
few-shot demonstrations, synthesis agents examine the issue,
repository, and merged PR to identify
implementation-critical information that is underspecified or implicit in the issue
and formulate candidate questions. Next, a validation agent checks each QA pair against the repository,
developer patch, and validated plan for evidential support, cross-stage
consistency, and absence of implementation detail leakage. Finally, two authors independently review every resulting
clarification reference for evidential support and cross-stage consistency.

\paragraph{Ground-Truth Construction for Planning.}
Planning-ground-truth construction has two phases: \textit{synthesis} and \textit{validation}.
During \textit{synthesis}, a single synthesis agent examines the original issue, base
repository, and merged PR, recovers the implementation intent, explains how the
developer patch realizes it, and organizes these findings into an ordered
sequence of modular steps. Each step specifies its goal, relevant files,
intended changes, and repository constraints. These fields make the step
actionable and auditable: they tell a coding agent what to achieve, where and
how to modify the repository, and which constraints to preserve, while enabling
step-level alignment with the gold patch.
During \textit{validation}, an alignment agent associates each plan step with the
corresponding subset of the complete developer patch. We call this subset the
step's \emph{gold subpatch}. We validate the plan sequentially, one step at a
time. For each step, a coding agent follows the step to generate a patch, and an
evaluator judges whether it is semantically equivalent to the corresponding
gold subpatch. If the check fails, a gap-analysis and resolution agent revises
the step plan. Steps that pass are retained.
This process establishes \emph{functional reproducibility}: an accepted plan
must provide sufficient guidance for a coding agent to produce a functionally
equivalent implementation. 
Finally, two authors independently review every resulting plan for evidential
support and cross-stage consistency.

\paragraph{Ground-Truth Construction for Code Generation.}
Unlike Req. Clarification and Planning, Code Generation requires no annotation
reconstruction. Each merged PR directly provides the developer implementation
patch, while its task-relevant tests serve as the executable correctness oracle.

\paragraph{Quality Assurance.}
As described above, we enforce quality control through the following
stage-specific validation and expert-review measures. At the task level, we
retain tasks with buildable environments and stable tests. For Planning, every step must
support an implementation functionally equivalent to its gold subpatch. For
Requirement Clarification, every point (QA pair) must be supported by repository
evidence, consistent with the validated plan and final implementation. 
Lastly, human experts review every retained task and
its GTs across all three stages.

\begin{figure*}[t]
  \centering
  \includegraphics[width=0.98\textwidth]{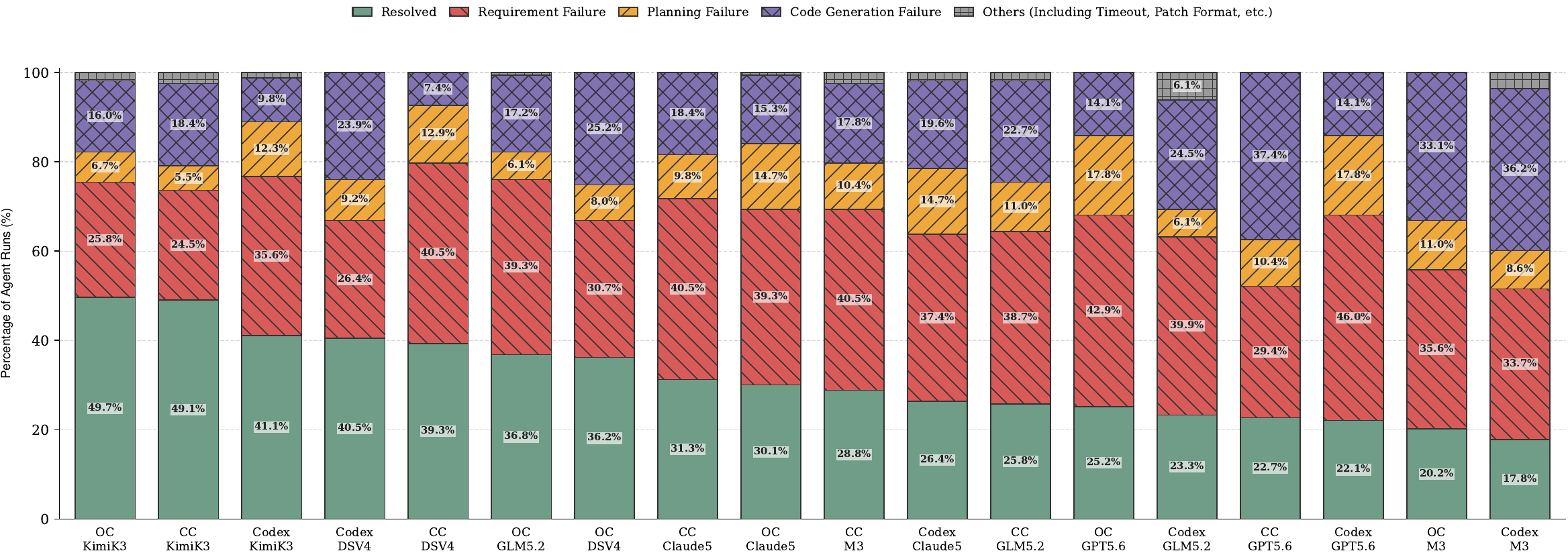}
  \caption{Failure diagnoses across agent--LLM configurations. Green denotes
  resolved runs; the remaining segments show the proportions attributed to
  requirement, planning, code-generation, and other failures (e.g., timeouts).}
  \label{fig:configuration-diagnostic-profiles}
  \end{figure*}

\subsection{Benchmark Statistics and Characteristics}
\label{sec:dataset-statistics}

Table~\ref{tab:construction-funnel} demonstrates the rigor and selectivity of
our quality-oriented construction: successive checks of task
relevance, environment reproducibility, test stability, and GT validity reduce
more than 2,000 real-world PR--issue candidates to 163 quality-controlled instances.
Table~\ref{tab:benchmark-gt-statistics} further reflects substantial
repository-scale complexity: codebases contain 272.51K lines of code (LOC) on
average, with the largest reaching 1.59M LOC; 
and each task
contains 2,248.56 functional test cases on average.
For each task, Clarification GT recovers 4.53 implicit and implementation-critical
requirement points on average, while Plan GT provides an average
of 2.06 actionable implementation steps with 11.80 implementation constraints.
Our benchmark covers across 7
engineering domains, including static analysis, compilers, and schema systems, 
algorithms, mathematics, and scientific computing.

\section{Experimental Setup}
\label{sec:experiments}

\paragraph{Evaluated Agents and Models.}
We evaluate three popular coding-agent frameworks (Claude Code, Codex, and
OpenCode) paired with six recent LLMs:
Claude-Sonnet-5~\citep{anthropic2026claudesonnet5},
DeepSeek-V4-Pro~\citep{deepseek2026v4pro}, GLM-5.2~\citep{zai2026glm52},
GPT-5.6-Terra~\citep{openai2026gpt56terra},
MiniMax-M3~\citep{minimax2026m3}, and
MoonshotAI-Kimi-K3~\citep{moonshot2026kimik3}.
We run each agent-LLM configuration twice
on every task and report run-level aggregates over the two runs.

\paragraph{Computing Infrastructure.}
All experiments were conducted on Ubuntu 26.04 LTS using an Intel Xeon 6982P-C
CPU (48 cores and 96 threads) with 182 GiB of RAM. We accessed all LLMs through
their official API providers.

\paragraph{Evaluation Metric.}
We use \textit{Resolve Rate} as the primary evaluation metric, following the
SWE-bench execution protocol~\citep{jimenez2024swebench}. A task is considered
resolved only if its patch applies successfully, all fail-to-pass tests pass,
and no pass-to-pass test regresses.

\paragraph{Failure Attribution to Stages.}
For each unresolved run, we use
\textit{GPT-5.6-Sol}~\citep{openai2026gpt56sol} as an LLM judge. Its dedicated
prompt jointly examines the task, intermediate-stage GTs, and the agent
trajectory. To make long trajectories tractable without discarding
stage-relevant evidence, we remove only verbose tool-call payloads and outputs
unrelated to requirement understanding, planning, implementation, or
verification, while retaining relevant repository observations, decisions,
edits, and validation actions. The judge assigns each failure to the earliest
deviating stage: requirement understanding, planning, implementation, or
others. On a stratified sample of 50 unresolved runs, its labels exactly match
human consensus in 46 cases (92\%).

\paragraph{Pre-code Stage Evaluation.}
Our Clarification and Plan GTs decompose each task into labeled information
points. Clarification spans six categories (C1--C6), whereas Planning covers
the goal, target location, implementation approach, constraints, and validation
strategy. For each run, the LLM judge (\textit{GPT-5.6-Sol}) determines whether the
agent's trajectory covers each GT information point. Because trajectories can
be long, we remove verbose tool-call payloads and outputs that carry no evidence
about requirement understanding or planning, while retaining relevant
repository observations, decisions, edits, and validation actions. We compute
category- and dimension-level coverage rates to evaluate the Clarification and Planning
stages from full-trajectory evidence rather than from the final patch outcome
alone. To accommodate multiple valid plans, the prompt evaluates semantic
implementation responsibilities rather than exact plan matching: it accepts
alternative files, symbols, architectures, and step orderings when they realize
the same responsibility and satisfy repository constraints, without requiring
textual or structural similarity to the reference plan. On a manually
annotated sample of 50 information-point assessments, the
judge's binary coverage decisions agree with the human consensus in 96\% of
cases.

\section{Results and Analysis}
\label{sec:results}

\subsection{Patch Correctness}

\paragraph{Overall Patch Correctness Results.}
On average, coding agents achieve a
resolved rate of only 31.5\%, indicating the difficulty of the \toolname{} tasks. 
At the configuration level, OpenCode with MoonshotAI-Kimi-K3 achieves the
highest resolved rate (49.7\%). Thus, even the best configuration leaves
roughly half of the tasks unresolved, indicating substantial room for
improvement in repository-level issue resolution.

\paragraph{Effectiveness of Agents, LLMs, and Pairings.}
Figure~\ref{fig:combo-heatmap} compares performance at the agent, LLM, and
agent--LLM pairing levels. Among agents, OpenCode achieves the highest average
resolved rate (33.0\%), followed closely by Claude Code (32.8\%) and Codex
(28.5\%). Among LLMs, MoonshotAI-Kimi-K3 performs best, averaging 46.6\% across
the three agents, followed by DeepSeek-V4-Pro at 38.7\%; the remaining backends
average 22.3--29.2\%. At the pairing level, OpenCode--MoonshotAI-Kimi-K3 is
strongest (49.7\%), closely followed by Claude Code--MoonshotAI-Kimi-K3
(49.1\%), whereas Codex--MiniMax-M3 is weakest (17.8\%). The resulting spread
shows substantial descriptive variation across backend--scaffold pairings.

\begin{figure*}[!t]
  \centering
  \includegraphics[width=0.95\textwidth]{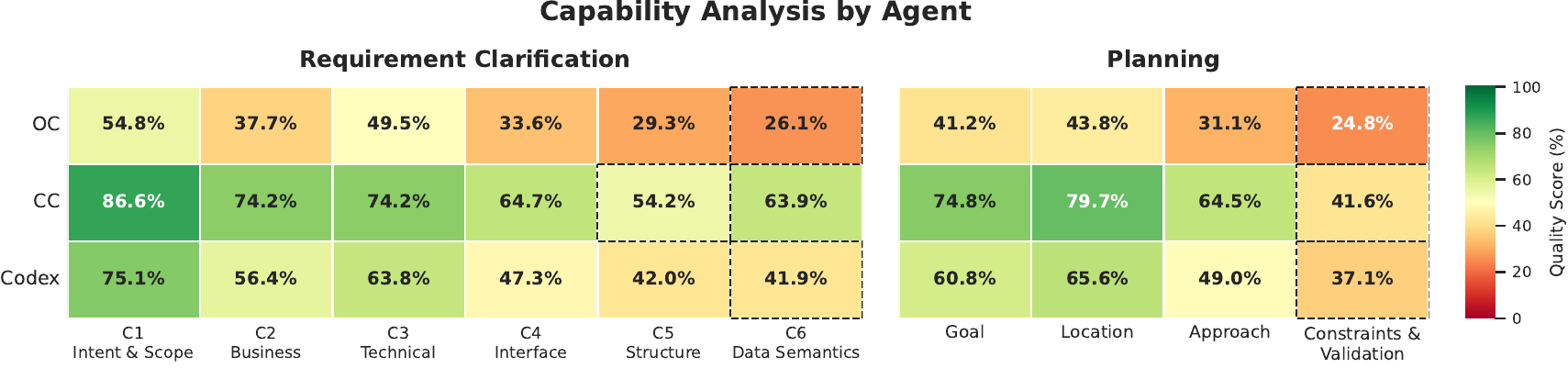}
  \caption{Clarification-category and planning-dimension coverage by agent,
  averaged across six LLMs. Boxes mark the lowest scores.}
  \label{fig:agent-capability-heatmap}
  \end{figure*}

\paragraph{Cost Analysis.}
Figure~\ref{fig:cost-time-resolve}(a) relates resolve rate to cost, with dashed
lines marking the average cost (\$1.59 per task) and average resolve rate
(31.5\%). OpenCode--MoonshotAI-Kimi-K3 achieves the highest resolve rate at
below-average cost, whereas Claude Code with the same backend attains similar
performance at above-average cost. DeepSeek-V4-Pro pairings likewise perform
above average but vary substantially in cost, showing that cost-effectiveness
depends on both the backend and agent scaffold.

\paragraph{Time Analysis.}
Figure~\ref{fig:cost-time-resolve}(b) shows a partial performance--time
trade-off. Many above-average configurations, particularly those using
MoonshotAI-Kimi-K3, exceed the average execution time of 8.5 minutes. However,
Claude Code--DeepSeek-V4-Pro performs above average in substantially less time,
while several slower configurations remain below average. Thus, longer
execution does not consistently produce better patches, and pairings exhibit
distinct time--performance trade-offs.

\subsection{Failure Diagnosis}
\label{sec:analysis-discussion}

Patch correctness reveals whether a run succeeds, but not why it fails. We
therefore align agent trajectories with validated Clarification and Plan GTs to
diagnose stage-specific failures. Figure~\ref{fig:configuration-diagnostic-profiles} 
decomposes results into resolved cases, and unresolved cases are attributed to main failure stages. 

\paragraph{Requirement failure is the largest component for most
configurations (24.5--46.0\%), followed by code-generation (7.4--37.4\%) and
planning (5.5--17.8\%) failures.}
This ordering identifies implicit requirement recovery as the most common bottleneck,
while planning errors are also non-negligible.
Code-generation failure rates show substantial descriptive variation across
LLM--scaffold pairings, particularly for several MiniMax-M3 pairings.

\paragraph{Similar resolve rates mask different bottleneck profiles.}
Codex-- and
Claude Code--DeepSeek-V4-Pro achieve similar resolve rates (40.5\% versus
39.3\%) but markedly different profiles: their requirement-failure rates are
26.4\% and 40.5\%, whereas their implementation/verification-failure rates are
23.9\% and 7.4\%. Similar end-to-end performance can therefore accompany
distinct stage-attribution profiles.

\paragraph{\toolname{} suggests candidate areas for agent improvement.}
Grounded in validated intermediate GTs, \toolname{} provides evidence
of where failures arise and identifies candidate modification areas beyond what
previous outcome-only evaluation can support. These diagnoses suggest
stage-specific directions: requirement failures motivate investigating stronger
implicit-constraint recovery, while planning failures motivate investigating
better implementation plans.

\subsection{Req. Clarification and Planning of Agents}
\label{sec:gt-diagnostic-granularity}
We compare each agent's trajectory with the validated Clarification and Plan GT
information points to measure how completely it captures the expected pre-code
information. Figure~\ref{fig:agent-capability-heatmap} reports coverage rates
by clarification category and planning dimension, averaged across 6 LLMs for each agent.

\paragraph{Lower clarification coverage is concentrated in interfaces,
structure, and data semantics.}
All three agents cover intent and scope most reliably, but weaken on the
implementation-facing categories. Claude Code's lowest clarification score is
code structure (54.2\%); Codex falls to 42.0\% on structure and 41.9\% on data
semantics; 
These results suggest candidate requirement-clarification targets, including
interface contracts, repository conventions, and data invariants beyond the
high-level intent.

\paragraph{Lower planning coverage is concentrated in approach,
constraints, and validation.}
For every agent, planning coverage decreases from target location to
implementation approach and then to constraints and validation: 79.7\% to
64.5\% to 41.6\% for Claude Code, 65.6\% to 49.0\% to 37.1\% for Codex, and
43.8\% to 31.1\% to 24.8\% for OpenCode. This consistent cascade shows that
agents often identify \emph{where} to edit without fully specifying \emph{how}
to implement and verify the change. These lower-coverage dimensions suggest
evaluating plans that more explicitly capture the implementation mechanism,
boundary conditions, compatibility requirements, and validation obligations.

\section{Conclusion}

\toolname{} combines 163 real PR-derived tasks, executable patch oracles, and
validated Clarification and Plan GTs. Beyond measuring final patch correctness,
it supports full-trajectory, GT-aligned retrospective diagnosis of requirement,
planning, implementation, and artifact gaps. These diagnostic profiles make
otherwise identical unresolved outcomes more interpretable and suggest
stage-specific hypotheses for future improvement.
The Code and Data Supplement contains the code needed to run the
benchmark and reproduce the reported analyses.

\section*{Limitations}

\toolname{} contains 163 Python and Java tasks from 31 repositories, so its
findings may not generalize to other languages or ecosystems.
Because each agent runs autonomously end to end, the benchmark cannot assess
the effect of human feedback or interactive refinement.
Its intermediate GTs and retrospective alignment labels depend on LLM-assisted
construction and judging, although all GTs are manually validated.
Future work should expand task and model coverage and strengthen annotation
audits, including inter-annotator agreement.
Generative AI tools assisted with manuscript drafting and language editing; the
authors reviewed and verified all content, claims, references, and final wording.

\bibliography{main}

\end{document}